\documentclass[12pt,preprint]{aastex}

\slugcomment{}

\shorttitle{SBS0335-052E Super Star Clusters}
\shortauthors{Thompson et al.}

\begin{document}

\title{Delayed Photoionization Feedback in a Super Star Cluster in SBS0335-052E}

\author{Rodger I. Thompson}
\affil{Steward Observatory, University of Arizona,
    Tucson, AZ 85721}
\email{rthompson@as.arizona.edu}

\author{Marc Sauvage}
\affil{Service d' Astrophysicque, CEA/DAPNIA, Centre d' Etudes de Saclay,}
\email{msauvage@cea.fr}

\author{Robert C. Kennicutt and Charles W. Engelbracht}
\affil{Steward Observatory, University of Arizona,
    Tucson, AZ 85721}
\email{rkennicutt@as.arizona.edu, cengelbracht@as.arizona.edu}

\author{Leonardo Vanzi}
\affil{European Southern Observatory, Alonzo de Cordova 3107,Santiago, Chile}
\email{lvanzi@eso.org}

\clearpage

\begin{abstract}

SBS0335-052 is a well studied Blue Compact Dwarf galaxy
with one of the lowest metallicities of any known galaxy.  It also 
contains 6 previously identified Super Star Clusters. We combine archival
HST NICMOS images in the Pa$\alpha$ line and the $1.6 \micron$
continuum of the eastern component, SBS0335-052E, with other space 
and ground based data to perform a multi-wavelength analysis of the
super star clusters.  We concentrate on the southern most clusters,
designated S1 and S2, which appear to be the youngest clusters and
are the strongest emitters of Pa$\alpha$, radio, and x-ray flux.
Our analysis leads to a possible model for S1 and perhaps S2 as a cluster of
very young, massive stars with strong stellar winds.  The wind
density can be high enough to absorb the majority of ionizing photons within
less than 1000 AU of the stars, creating very compact HII regions
that emit optically thick radiation at radio wavelengths.  These winds would
then effectively quench the photoionizing flux very close to the stars. This
can delay the onset of negative feedback by photoionization and 
photodissociation on star formation in the clusters.  This is significant
since SBS0335-052E resembles the conditions that were probably common 
for high redshift star formation in galaxies near the epoch of reionization. 

\end{abstract}


\section{Introduction} \label{s-i}

\begin{figure}
\plottwo{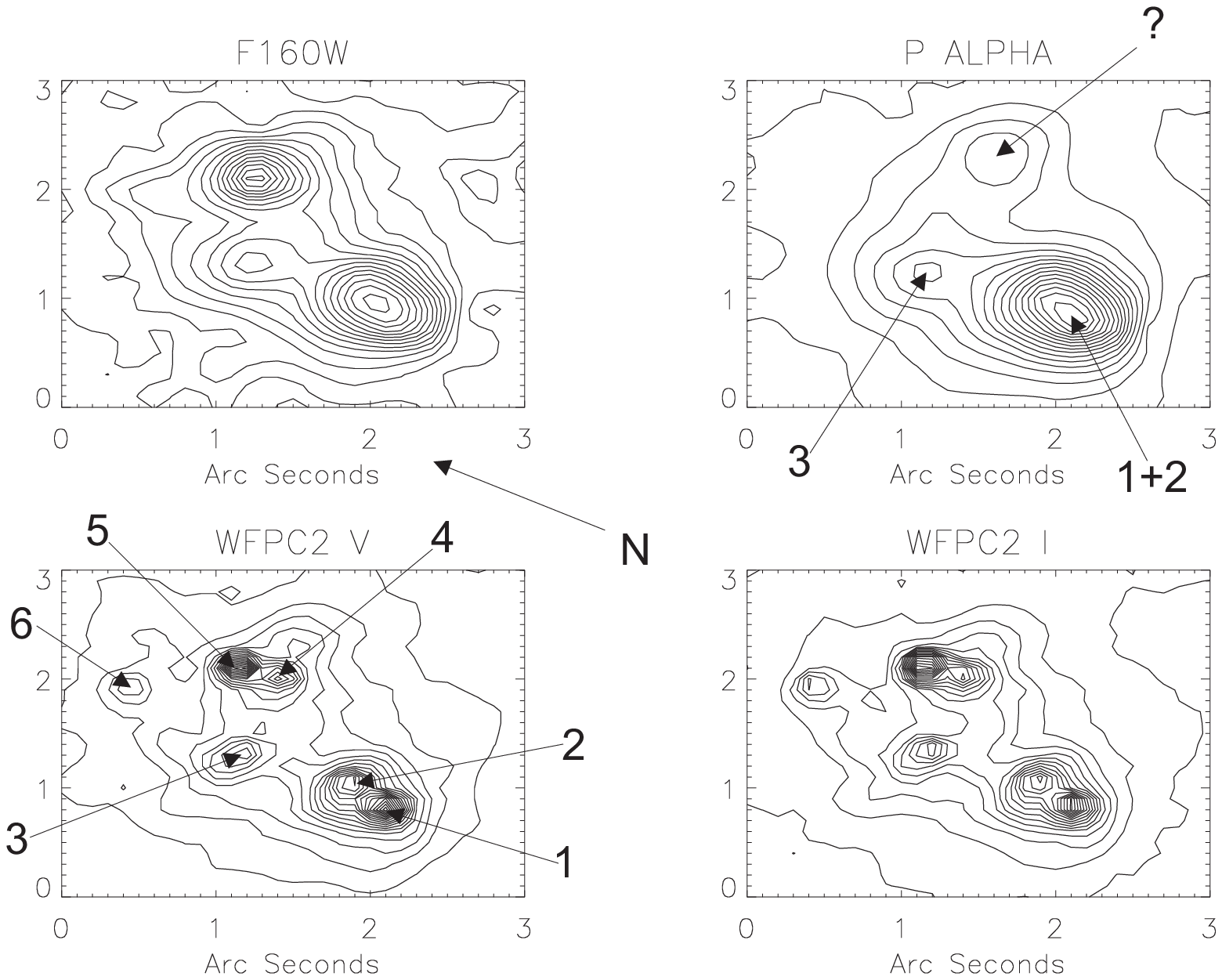}{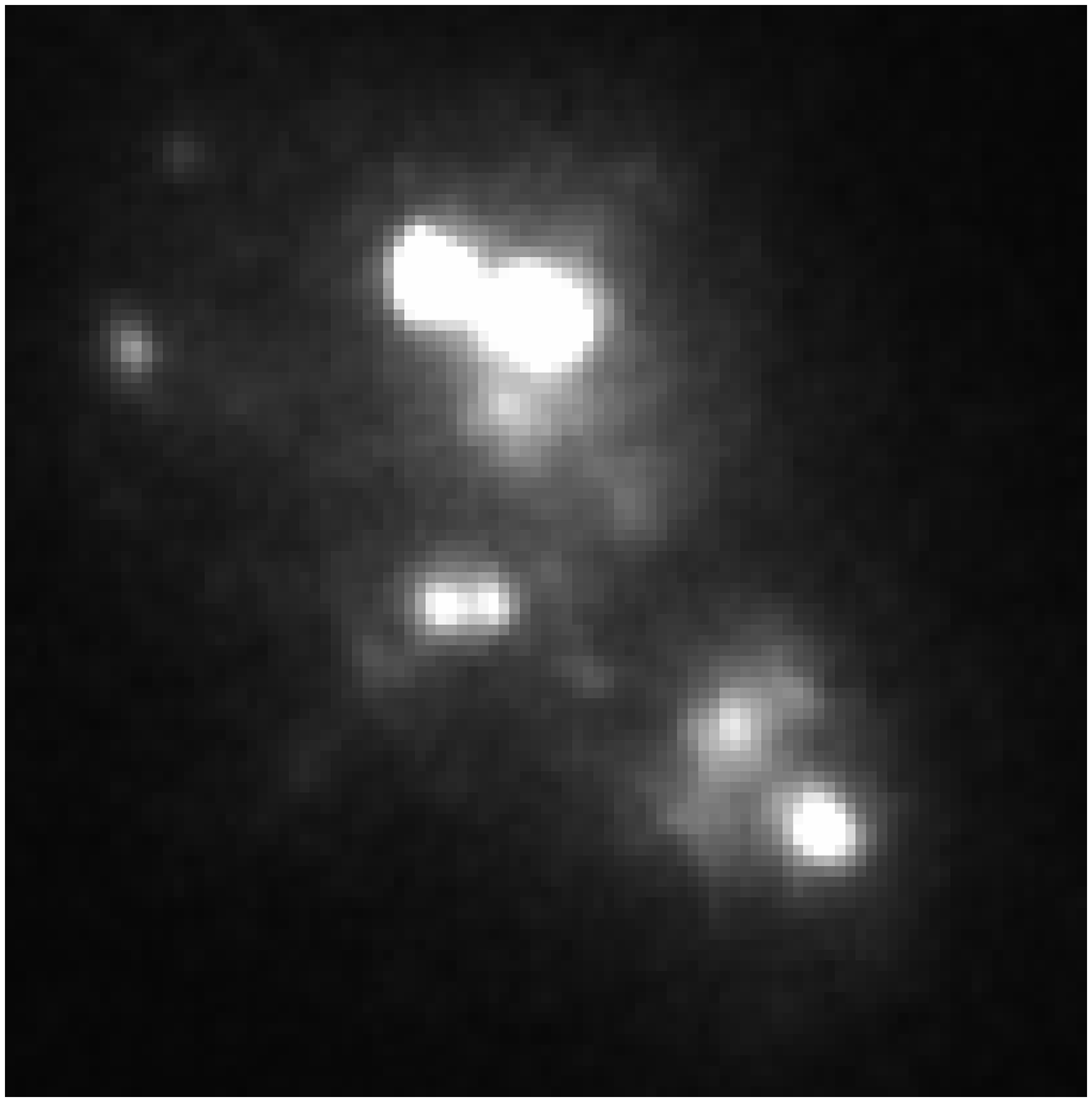}
\caption{a) Contour plots of the SBS0335-052E sources from WFPC2 F569W (V) 
and F791W (I) images compared with the NICMOS F160W and Pa$\alpha$ camera
3 images.  The source numbering in the V image is from \citet{thu97} and
the source orientation is the same as in that work. North is indicated 
by the arrow. The source marked ? in the Pa$\alpha$ image is a line
emission peak with no corresponding continuum image. b) On the right is
the ACS continuum image (F149LP) at the same rotation but enlarged.
As noted in \citet{kun03} the source labeled 3 is actually two sources
at the ACS resolution. \label{fig-src}}
\end{figure}

Many present day simulations of galaxy formation predict a much larger
fraction of baryons that eventually become stars than the observed 
fraction of 5 to 10\% 
\citep{col01}.  At the baryon densities favored by
CMB measurements the luminosity function is over predicted at both
high and low luminosities \citep{ben03}. Introduction of negative feedback by 
supernovae has not eliminated the problem.  Negative feedback from the
processes of star formation may offer the solution. SBS0335-052E, 
found in the Second Byurakan Survey \citep{mar83}, offers a unique 
opportunity to observe star formation and feedback in conditions similar
to primordial star formation but in a relatively nearby (53 Mpc) galaxy.
SBS0335-052E resembles our current expectations for very early galaxy
formation in that it has a very low metallicity, Z$_{\sun}$/40 (\citet{iz90},
\citet{iz92})
and a very high gas to stellar mass ratio, $\sim 25$ to 1, based on
the stellar content visible in optical light \citep{pus01,pap98}. 
Observations with WFPC2 on HST by \citet{thu97} revealed 6 point like
sources, (see fig.~\ref{fig-src}a for the source numbers and contour
plots) superimposed on diffuse emission.  They deemed the 6 sources,
which are within 500 pc of each other, to be Super Star Clusters (SSCs)
with current star formation.  

ISO observations by \citet{thu99} found strong dust emission from the 
region of the sources. These observations were supplemented by 
\citet{dal01} at 12.5 $\micron$ and other wavelengths and by 
\citet{hun01} at 4 $\micron$. Recent Spitzer observations \citep{hou04} 
measured the spectrum of the infrared emission and found it quite different 
than other starburst galaxies and devoid of any PAH emission features
as was noted by \citet{thu99}. 
Additional Spitzer IRAC and MIPS data have also been recently acquired
by \citet{eng05}. Non-thermal radio emission with a possible optically 
thick thermal component \citep{hun04} has been found to come from the 
region of the sources S1 and S2. The resolution of the radio observations is not 
high enough to spatially separate the two sources, however, very recent
radio observations by \citet{joh05} resolve the two sources. Sources S1 
and S2 are the southern most sources of the 6 SSCs which are bounded in 
the north by a supernova bubble which has led to speculation of a sequence 
of star formation propagating from north to south \citep{thu97}.

Chandra x-ray observations \citep{thu04} detect a point source at a
position consistent with source 2 and possibly source 1.  \citet{thu04}
conclude that the source is consistent with a compact group of high
mass x-ray binaries or an intermediate mass black hole.   
\citet{pus04} have analyzed the optical colors and H$\alpha$ emission
in SBS0335-052E to determine the maximum age of the underlying Low
Surface Brightness (LSB) component of the galaxy. They find 
ages of $\lesssim(100, 400)$ Myr for (instantaneous, constant) star
formation models with the instantaneous model providing the best
fit, however previous work by \citet{ost01} has provided caveats
to the age determination in SBS0335-052.

\begin{figure}
\plotone{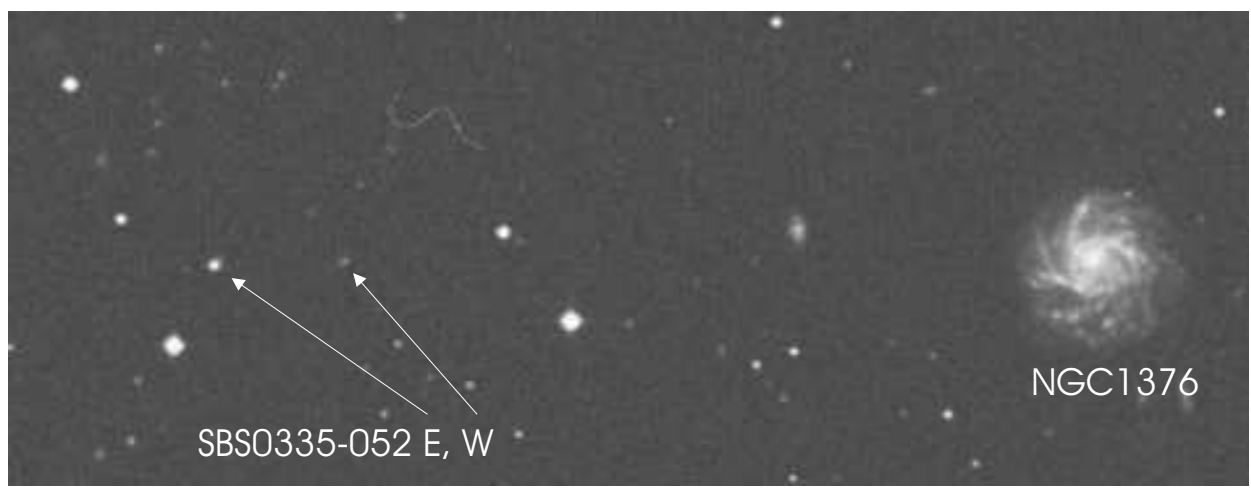}
\caption{SBS0335-052 and NGC1376.  The SSCs discussed here lie in 
SBS0335-052E and are completely unresolved at the spatial scale of this
image.  The image is taken from the DSS image prepared by STScI from
images obtained using the Oschin Schmidt Telescope on Palomar Mountain.
 \label{fig-reg}}
\end{figure}

The SSCs of SBS0335-052E lie in the eastern component of the 
SBS0335-052E,W complex.  Radio HI observations by \citet{pus01}
find that both components are part of a large neutral hydrogen
cloud of about $2 \times 10^9 M_{\sun}$ with a significantly larger
spatial extent than most BCDs.   \citet{pus01} also determine a
dynamical mass of $9 \times 10^9 M_{\sun}$ for the system, most
of which must be dark matter.  They also note the presence of
NGC 1376 at a projected distance of 150 kpc with a radial velocity
about 100 km/sec greater than the SBS0335-052E system and a mass of
0.5 to $1.0 \times 10^{12} M_{\sun}$. Figure~\ref{fig-reg} shows
the spatial relationship of NGC 1376 and SBS0335-052 E and W.
Interaction with NGC 1376 may be the trigger that initiated the
current star formation episode in the SSCs \citep{pus01} and in
SBS0335-052W which new work by \citet{izo05} has found to have
the lowest metallicity yet found for any galaxy.

The first HST UV spectra were presented by \citet{thu97b} which
showed Lyman alpha in absorption, consistent with the large neutral
hydrogen cloud observed at radio wavelengths. They also found UV metal
lines with P Cygni profiles which are most likely associated with
the UV bright sources S4 and S5. HST ACS UV images of SBS0335-052E,
fig.~\ref{fig-src}b ,using the solar blind channel by \citet{kun03} 
found very little, if any, Ly$\alpha$ emission escaping from the SSCs. 
They also revealed that S3 is double with a separation of 
only 26 pc projected on the sky.  HST UV spectra 
with STIS \citep{kun04} also show strong absorption in Ly$\alpha$ 
consistent with previous observations. \citet{thu97b} calculate a
column density of neutral hydrogen of $7.0 \times 10^{21}$ cm$^{-2}$,
again consistent with the large HI mass determined from the radio
observations. UV spectroscopy with FUSE \citep{thu05} confirms this
result and measures a low metal abundance in the neutral gas which
is similar to the ionized gas in contrast with earlier results 
\citep{thu97b}. They note that their spectrum  does not show any
H$_2$ absorption lines. The near infrared H$_2$ emission spectrum observed 
by \citet{van00} contains both thermal shock and florescent excitation
components indicating a mixture of the two excitation mechanisms. 
Since the FUSE observations encompass the entire set of SSCs
the primary line of sight probed by the observations is that toward
S4 and S5 which are the brightest sources in the UV continuum.
The neutral gas metal measurements are therefore dominated by
the gas along those sight lines.  If the H$_2$ emission 
is primarily associated with the more obscured
S1 and S2, the lack of UV H$_2$ absorption lines quoted by
\citet{thu05} can easily be explained.  

The general picture of SBS0335-052E is a low metallicity, gas rich 
system with ongoing star formation in SSCs.  The bulk of the underlying 
stellar population appears to be less than at most 400 Myr indicating that 
most of the stars in SBS0335-052E are the result of recent star formation.  
The metals that do exist in SBS0335-052E may be a remnant of a much earlier 
episode of star formation or the result of an initial seeding of the
IGM from the Pop. III stars that reionized the universe. We examine 
SBS0335-052 as a local example
of star formation processes that must have been common in the early
universe. If SBS0335-052 eventually becomes part of NGC 1376 it will
be an example of a minor merger in the hierarchical process of galaxy
assembly. In the following we will concentrate on the southern most
sources, S1 and S2, which appear to be the youngest of the SSCs and
the major source of the x-ray, Pa $\alpha$, and radio emission.  The
Pa $\alpha$ and radio fluxes, plus the ground based determined near IR
extinction are the keys to building a consistent picture of S1 and S2
and judging the feedback effect of star formation in the sources.  In
this paper the primary question is not the age of the underlying stellar
population but rather what is the effect of the current star formation
on the immediate environment.

\section{Observations} \label{s-obs}

The archival data analyzed here are from HST NICMOS images in
F160W and Pa $\alpha$ (GO-9360, R. Kennicut PI) which were taken as a component of
the SIRTF (now Spitzer) Infrared Nearby Galaxies Survey (SINGS) \citep{ken03}.  
All images are from the NICMOS camera 3 which has 0.2 arc 
second pixels for a wide non-diffraction limited field.
The observations occurred on July 15, 2002.
There are (4) 23.97 second integrations in the F160W filter, (4)
191.96 second integrations in the F187N filter and (4) 159.96
second integrations in the F190N filter.  This is the standard
integration suite for the SINGS snapshot survey.  The redshift
of SBS0335-052E puts the Pa $\alpha$ line into the F190N filter
rather than the rest frame F187N Pa $\alpha$ filter.  In this case
the F187N filter serves as the continuum filter.  The narrow
band integrations were both STEP32 with 14 reads in the F187N
filter and 13 reads in the F190N filter.  The F160W integrations
were STEP8 with 9 reads.  The observations used a 4 point spiral
dither pattern with $0.9\arcsec$ spacing between the points. 
Integrations in each of the three filters were made at each
dither position before moving to the next position.  The
integrations were carried out in a single orbit.

\section{Data Reduction} \label{s-dr}

The NICMOS images were reduced in a standard manner (see 
\citet{thm05} for a detailed description).  Although 
auxiliary observations were available for correction of
SAA passage persistence, they were not utilized as none
of the images appeared to have suffered from the effect.
The NICMOS data were then rebinned to the WFPC2 pixel size
of $0.1\arcsec$ with the procedure IDP3 \citep{lyt99}.
The images were aligned utilizing stars in the image and
the F187N image was subtracted from the F190N image to
produce the Pa $\alpha$ image.  The NICMOS image was then
aligned to the original WFPC2 images of \citet{thu97}
again using stars in the image.  Initially this was
done using only the brightest star in the image, however,
this appeared to produce poor alignments.  It was then
discovered by aligning using the fainter stars in the field
that the brightest star has significant proper motion and
had moved by $0.18\arcsec$ between the NICMOS observations
and the WFPC2 images taken in January of 1995.  The alignment
accuracy is better than $0.03\arcsec$. Contour plots of the
NICMOS and WFPC images are shown in fig.~\ref{fig-src}a.

The measured flux in Pa $\alpha$ for S1 and S2 is $1.77 \times
10^{-14}$ ergs s$^{-1}$ cm$^{-2}$.  The point source limit
for detection is about a factor of 100 below this number but
varies across the image depending on the amount of diffuse
emission in the region. The 1 $\sigma$ single pixel noise is
0.002 ADUs per second which is equivalent to $1.2 \times 10^{-18}$
ergs s$^{-1}$ cm$^{-2}$.  The continuum subtraction is accurate 
to better than $5\%$ based on the net residuals in the areas
of the image that are devoid os sources.  The F160W image has
a 1 $\sigma$ single pixel noise of 0.01 ADUs per second while
the S1+S2 source has a total signal of 62 ADUs per second.

We also retrieved the archival ACS UV images from Proposal 9470
with Daniel Kunth as PI which are described in \citet{kun03}.
These were aligned with IDP3 and utilized to better understand
the morphology of SBS0335-052E at higher spatial resolution than
the NICMOS images or the previous WFPC2 images.  These images are
in the continuum F140LP and Ly$\alpha$ F122M filters.  The pixel
size of the retrieved images is 0.025\arcsec which were subsequently
rebinned to 0.1\arcsec to match the other images used in this
investigation.  The single pixel 1 $\sigma$ noise of the F140LP
image is $4.0 \times 10^{-21}$ ergs s$^{-1}$ cm$^{-2}$.  Since
Ly$\alpha$ is in absorption it is unclear how to interpret the
F122M image, therefore the F140LP image is the only ACS image used 
in this study.  Using a point spread function (PSF) determined from
stars in the UV image away from the SSCs we find that the SSC S1 is
resolved with a radius of 3 pc assuming the 53 Mpc distance given in
\S~\ref{s-i}. 

\section{The Volume Emission Measure} \label{s-vem}

The key elements to this study are the volume emission measure,
$N_e^2V$, and the nature of the thermal radio emission.  Together
they determine the nature of the stellar population and the state
of the ionized gas in S1 and S2.  In what follows we will lump S1
and S2 together since they are unresolved in the NICMOS
image.  We assign half of the emission to each source,
and hence half of the volume emission measure,
although a radio image from \citet{joh05} indicates that S1 is
the stronger radio source. The error is at most a factor of two 
and does not affect the major conclusions of the paper. The optical and 
near infrared contour plots and the UV F149LP image are shown in 
Figure~\ref{fig-src} in the orientation of the original WFPC2 images. 
The dominant Pa $\alpha$ emission is clearly centered on S1 and S2
with some emission also at S3.  The lack of emission at the other
sources may be due to lack of integration time.  The gradient in
Pa $\alpha$ emission is consistent with the propagation of star
formation from the north to the south as discussed in \S~\ref{s-i}.

We next use the Pa $\alpha$ image to determine the volume emission
measure for S1 and S2 which in turn is a direct measure of
the ionizing flux. \citet{hun01} measured the Br $\gamma$ and
Br $\alpha$ flux in a 1.5 by 1 arc second region centered on
S1 and S2 that does not include any of the emission from S3.  The 
measured Pa $\alpha$ flux in this region is $1.77 \times 10^{-14}$ 
ergs s$^{-1}$ cm$^{-2}$ after subtraction of flux from an aperture
of equal area off the source to eliminate any contribution due to
diffuse emission. The observed $Br \gamma / Br \alpha$ ratio \citep{hun01}
predicts an A$_V$ value of $12.1 \pm 2$ using the extinction law 
of \citet{rie85} which is very similar to the value of 
12.2 found by \citet{hun01} using another extinction law.  This in 
turn predicts an extinction of 1.64 magnitudes at Pa $\alpha$ which
gives an extinction corrected Pa $\alpha$ flux of $8.01 \times 10^{-14}$ 
ergs s$^{-1}$ cm$^{-2}$ for the 1 by 1.5 arc second region that contains S1 and S2. 
Here we have assumed that since S1+S2 is the brightest Pa $\alpha$
emitter, it is also probably the brightest Br $\alpha$ emitter
as well and the measured extinction from $Br \gamma / Br \alpha$ is
appropriate for correcting the Pa $\alpha$ emission of S1+S2.
\citet{izo97} find that the optical emission lines are best
matched with an ionized gas temperature of 20,000 K which we
will use in the rest of the analysis.  An important question
is whether the observed Pa $\alpha$ flux in this work is
consistent with the Br $\alpha$ measurements in the context
of normal Case B recombination theory. The extinction at 
Br $\alpha$ from \citet{rie85} is 0.57 magnitudes which gives an
extinction corrected Br $\alpha$ flux of $1.52 \times 10^{-4}$
ergs s$^{-1}$ and a Pa $\alpha$ to Br $\alpha$ ratio of 5.26.
The line ratios given in the appendix of \citet{hun04} show that
the calculated ratio for 20,000 K is 4.22.  Given the expected errors
in fluxes and extinctions this is considered in reasonable agreement.
Note that if the Pa $\alpha$ to Br $\alpha$ ratio were affected by line
optical depth effects we would expect that the observed ratio would
be less than the calculated ratio since the population of the n=3 
state of the hydrogen gas should be greater than the n=4 state. The
numbers in this calculation are listed in Table~\ref{tbl-paba}.

A caveat to the calculated extinction should be mentioned.  If the extinction
derived from the infrared lines is extrapolated to the ultraviolet wavelength
of the ACS image, 1400 \AA it should be on the order of 24 magnitudes.  Comparison
of the predicted flux at 1400 \AA from the Starburst99 calculation described
below indicates an extinction of approximately 8 magnitudes.  This however
follows an already observed trend in SBS0335-052E that the derived extinction
decreases at shorter wavelengths \citep{hun01}.  The usual explanation is 
that only at longer wavelengths can you see into the regions of higher
extinction.  Since we are concerned with the true Pa $\alpha$ flux we will
use the extiction dervived directly from the infrared lines which is consistent
in the ratios of Pa $\alpha$ Br $\gamma$ and Br $\alpha$.  The radio images
of \citet{joh05} (slide 10) confirm that there are no hidden ionization regions off the
observed sources that could be the source of additional Br $\alpha$ flux that
might contaminate the measurement.

Using case B recombination theory 

\begin{equation}
\int_{\nu_0}^{\infty}\frac{L_{\nu}}{h\nu} = Q(H^0) = N_e^2V\alpha_B
\label{eq-strom}
\end{equation}

\noindent and the appendix from \citet{hun04}
the volume emission measure to total Pa $\alpha$ power is given by

\begin{equation}
N_e^2V = \frac{F_{Pa \alpha} (\frac{T}{10^4K})^{1.161}}{4.19\times10^{-26}}
\label{eq-ne2v}
\end{equation}

\noindent where $F_{Pa \alpha}$ is in ergs per second.  For a 
20,000 K ionized gas the observed Pa $\alpha$ for S1 only, assigning
half of the S1+S2 emission measure to S1, requires a volume 
emission measure (N$_e^2V$) of $7.2 \times 10^{65}$ cm$^{-3}$ 
which is equivalent to 16,000 O7 stars using the O7 star N$_e^2V$ 
value of $4.43 \times 10^{61}$ from \citet{str03}.  Use of the
standard \citet{ken98} relation between the number of ionizing
photons and the current star formation rate gives a rate of
2 M$_\sun$ per year for S1. Our concentration on S1 here, rather
than S1+S2 is slightly arbitrary since the Pa $\alpha$ image does
not resolve the two sources. However, as will be seen later, 
S1 has a resolved radio spectrum, which is not available yet for
S2, and S1 has a resolved size from the ACS UV image. S2 in this
image is a diffuse object.  We therefore concentrate on S1 as the 
source with the best determined physical parameters, and as mentioned
earlier, changes by a factor of 2 in the S1 emission measure does
not affect the conclusions of this paper. 

Although the absolute star formation rate of 2 M$_{\sun}$ per year 
for S1 is not particularly high, the specific star formation rate,
SSFR, the star formation rate divided by the stellar mass of 
$7.5 \times 10^6$ M$_{\sun}$ found below, is $1.4 \times 10^{-7}$ 
year$^{-1}$, consistent with the enhanced SSFR of
higher z galaxies from the plots in \citet{bau05}.  They find that 
the SSFR increases with redshift out to their maximum redshift of 1.5 
indicating higher star formation rates per unit mass by two orders of 
magnitude at z = 1.5 than today. This provides some evidence that the
star formation in SBS0335-052 is similar to star formation at much earlier
times.

We used the online STARBURST99 code \citep{lei99} to determine
the volume emission measure produced by a $10^6$ M$_{\sun}$ stellar 
population with a Salpeter IMF in the mass range between 0.1 and 
120 M$_{\sun}$. Comparing the volume emission measure from the 
STARBURST99 output at 3 My to the measured value 
we determine that S1 contains a stellar mass of $7.5 \times 10^6$ 
M$_{\sun}$. Stars below 1 M$_{\sun}$ are probably not yet on the main 
sequence but we assume that they are gravitationally bound protostars 
that contribute to the gravitational mass of the cluster.  Counting
only stars above 1 M$_{\sun}$, the total formed mass is approximately 
$3 \times 10^6$ M$_{\sun}$. The output of this same calculation determined 
that the actual number of O9 and earlier stars is approximately 7,500.  
This is lower than the number of equivalent O7 stars since the presence 
of stars earlier than O7 increases the number of ionizing photons per unit 
mass.  The total luminosity of the stellar population is $4.3 \times 10^9$ 
L$_{\sun}$ which is approximately 3 times the luminosity of the 
mid-infrared emission discussed in \S~\ref{sss-ir}. 

\subsection{Line Excess Phenomenon} \label{ss-le}

Observations of the Br $\gamma$ line in obscured Young Stellar Objects
(YSOs) in the 1980s indicated that the observed line flux for objects
with luminosities less than that of an O9 Zero Age Main Sequence (ZAMS) 
star were higher than predicted from the Ly continuum
emission expected from the star, eg. (\citep{thm82}, \citep{thm84}). At 
the same time the radio emission was far less than that from
optically thin free-free emission with the same emission measure.
The explanation of the line excess phenomenon was provided by
\citet{sim83} who showed that a star with a significant wind will
have a high density region which collisionally excites the n = 2
level of hydrogen. The excited atoms can then be ionized by Balmer
continuum photons which gives an excess line emission over the 
ionization expected from only Lyman continuum photons. N$_e^2V$
as derived from the line emission is then no longer an accurate measure
of the number of Lyman continuum photons.  The lack of radio emission
was due to the high optical depth at radio frequencies. If the majority 
of ionizing photons are produced by stars earlier than O9 then the line 
excess effect is irrelevant in our calculations involving N$_e^2V$.  
Calculations of the ionizing flux for a stellar population with a 
Salpeter IMF and an upper mass cutoff of 120 M$_\sun$ indicates that 98\% 
of the ionizing flux comes from stars earlier than O9, therefore the
calculated value of N$_e^2V$ is suitable for determining the nature
of the stellar population.

\section{Radio Component} \label{s-rc}

The optically thin free-free emission is directly proportional
to the volume emission measure by the thermal bremsstrahlung equation

\begin{equation}
\frac{dP}{d\nu} = 6.8 \times 10^{-51} T^{-1/2}N_e^2V g(\nu,T) exp(-\frac{h\nu}{kT})
\label{eq-ff}
\end{equation}

\noindent where $\frac{dP}{d\nu}$ is in W m$^{-3}$ Hz$^{-1}$ and all units
are in the mks system \citep{wo00}. For radio frequencies the 
Gaunt factor $g(\nu,T)$ is given by

\begin{equation}
g(\nu,T) = 0.28[ln(4.4 \times 10^{16} T^3 \nu^{-2})-0.76]
\label{eq-gau}
\end{equation}

\noindent As in Eqn.~\ref{eq-ff} the Z dependence has been dropped 
since the majority of the gas is hydrogen. At a distance of 53 Mpc 
the value of the volume emission measure found in \S~\ref{s-vem} from 
the Pa $\alpha$ flux of S1, converted to m$^{-3}$, predicts a thermal radio 
emission of 0.61 mJy at 10 GHz for an ionized gas temperature of 20,000 K
which is a factor of 4 above the flux observed by \citet{joh05} (slide 8) of 
0.15 mJy at 10 GHz for S1 only. The synchrotron flux observed by 
\citet{hun04} for S1 + S2 is not present in the \citet{joh05} spectrum,
indicating that it is probably associated with S2 or other regions 
outside S1. 

Instead of the expected $\nu^{-0.1}$ decrease in radio flux with frequency
spectrum of optically thin free-free emission the observed radio spectrum
has a roughly $\nu^{0.28}$ rising spectrum.  This, coupled with the
factor of 4 less emission than predicted from the emission measure,
indicates that the radio emission has a significant optically thick 
component.  The radio emission probably comes from a complex geometry
of varying optical depths and perhaps temperatures but observations of
galactic HII regions which surround the ionizing star \citep{sim83} suggests a very
simplified model of 7500 O stars with winds that produce HII regions
of high enough density that the Stromgren spheres are confined to areas
close to the star.  We can use the simplified models of \citet{oln75}
as a guide.  

\citet{oln75} showed that for a density structure of $(\frac{r}{R})^{-q}$
around a star the frequency dependence of the optically thick radio
emission is $\nu^n$ where $n = \frac{2q-3.1}{q-0.5}$.  The unhindered
free expansion wind with $q=2$ gives a roughly $n=\frac{2}{3}$ spectrum.
The value of $q$ for $n=0.28$ is 1.72 which is appropriate for a wind that
is not allowed to freely expand but is hindered by the surrounding medium.
The flux at a frequency $\nu$ is given by

\begin{equation}
S_{\nu} = \frac{2 \pi k T}{c^2 D^2} R^2 \nu^2 (\tau_{\nu}(R))^{\frac{2}{2q-1}} \Gamma(\frac{2q-3}{2q-1})
\label{eq-snu}
\end{equation}

\noindent where D is the distance to the object and R is the R in the 
density equation.  The optical depth $\tau_{\nu}(R)$ is given by

\begin{equation}
\tau_{\nu}(R) = (n_e)^2 \sqrt{\pi} R f(\nu,T) {\Gamma(q-1/2)}/{\Gamma(q)}
\label{eq-opt}
\end{equation}

\noindent and the emissivity f in $cm^6pc$ is given by

\begin{equation}
f(\nu,T) = 8.235 \times 10^{-2} (\frac{T}{k})^{-1.35} (\frac{\nu}{GHz})^{-2.1}
\label{eq-em}
\end{equation}

\noindent for Olnon's Model IV.

Distributing the observed flux equally among 7,500 stars and matching the
observed flux at 10 GHz gives the following parameters for the HII regions
around each star. The electron 
density at R is $1.6 \times 10^6$ and R is $2.1 \times 10^{-3}$ pc or
433 AU for a temperature of 20,000 K.  The optical depth at R is 9.7,
justifying the assumption of optically thick emission.  As is well known,
Olnon's Model IV is not valid for $q \leq 2$ since an integration of the 
density law leads to an infinite mass.  The density law must be truncated
at some radius as is done for $q=2$ in Olnon's model V.  The solutions are
not given for other q values but the general nature is that they are
basically Model IV in the optically thick regions and the normal free-free
spectrum in the optically thin regions, therefore use of Model IV in this
frequency range is justified and real stars will have truncated flows when
the pressure of the interstellar medium halts the wind.  The parameters 
described above are typical of hyper compact HII regions.  If 2R is taken as
a characteristic size of the ionized regions then the total volume occupied
by the ionized gas is only $4.6 \times 10^{-3}$ pc$^3$ which is a filling
factor of less than $10^{-4}$ for a 3 pc cluster radius.  This is justification
for our assumption that the HII regions are confined to the stars themselves.
The average distance between the 15,000 O stars in the cluster is about 0.1 pc,
much greater than the size of the HII regions. Note that integration of the
density equation out to 1.9 R accounts for the volume emission measure calculated
from the Pa $\alpha$ emission.  In reality it is probable
that although the majority of O stars may have confined HII regions, that some
of the HII regions have broken out and that some ionizing UV emission is
present throughout the cluster, producing a background optically thin 
component of emission. Obviously the stars in S1 will have differing density
structures and the structure we have used here is a representative average. 
The variations in structure will produce bumps and wiggles in the radio spectrum
as is observed in the spectrum of \citet{joh05}. It also should be emphasized that 
the model developed here is
not the only possible geometry of the ionized gas that will account
for the line and radio emission.  It is, however, motivated by observations
in our own galaxy of young stars with winds that limit the ionized area
to a small region centered on the star.

\subsection{Other Emission Components} \label{ss-oth}

Although not critical to our discussion of feedback in SBS0335-052E, we 
address two other known emission components for completeness and to show
that neither component has a significant effect on the galaxy at the 
present time.  The x-ray emission component could conceivably have a 
significant role in the nature of the galaxy at a later time, however.

\subsubsection{X-Ray Emission} \label{sss-xr}

The total x-ray luminosity for SBS0335-052E is approximately
$3.0 \pm .5 \times 10^{39}$ ergs per second \citep{thu04}
depending on the spectral model used to fit the data.  The
emission is primarily from a point source although \citet{thu04}
state that there is also evidence for faint extended emission
with a luminosity of $6.4 \times 10^{37}$ ergs per second.
The temperature of the x-ray emitting gas is $kT = 2.7_{-1.3}^{16.6}$
keV which is much hotter than the ~1 eV gas responsible for
the nebular continuum.  For this temperature gas the volume
emission measure, $N_e^2V$, required is $3 \times 10^{62}$ 
cm$^{-3}$ which is more than 1000 times less than is required
for the nebular emission (see \S~\ref{s-vem}).  This indicates
that the hot x-ray emitting gas is a minor component of the
ionized gas in SBS0335-052E. \citet{thu04} indicate that the
x-ray spectrum is relatively soft and consistent with either
a cluster of high mass x-ray binaries or with an intermediate
mass black hole.  The thermal bremsstrahlung from the x-ray 
emitting gas is only $10^{-8}$ Jy at radio frequencies, therefore 
it does not contribute to the observed radio flux. 

\subsubsection{Mid-Infrared Emission} \label{sss-ir}

One of the surprises of SBS0335-052E was the discovery of strong 
mid-infrared emission from dust in a galaxy that has very low
metallicity (\citet{thu99}, \citet{pla02}).  Neither the ISO image 
nor the SPITZER peak up image \citep{hou04} have the spatial resolution to 
associate the infrared emission with a particular component or SSC,
but it is consistent with being centered on S1-S2. The simplest explanation 
is that infrared source is associated with S1 and S2. This is also
consistent with the ground based observations of \citet{dal01} which
show that the Br$\gamma$ emission and the 12.5 $\micron$ emission are
coincident to an accuracy of 0.5\arcsec. The high emission measure of  
S1 and S2 indicate that they are young and still have dust nearby
if not in the clusters themselves. \citet{hou04} comment that
the mid-ir spectrum is consistent with very hot dust which is 
another reason for associating the mid-ir emission with the young
S1 and S2 SSCs. The intrinsic luminosity of
$4.3 \times 10^9 L_{\sun}$ is sufficient to power the approximately
$1.5 \times 10^9 L_{\sun}$ \citep{eng05} mid infrared luminosity. In fact
with an A$_V$ value of 12 there should be a three times higher mid-infrared
luminosity than that observed if the dust causing the extinction is located
close to S1.  Given our model of sources with strong winds it is probable
that the dust geometry is quite complex and that there may not be high dust
obscuration over a full $4\pi$ steradians, allowing a large fraction of luminosity
to escape in directions other than along our line of sight to the sources.  It
is very probable that there is little, if any, dust in the strong wind, 
ultra-compact HII regions surrounding the stars.

\section{Feedback on the Galaxy}

One of the critical missing elements in our ability to model the formation,
growth, and evolution of galaxies is a good understanding of how the 
initial star formation affects, or feedbacks on, subsequent star formation.
SBS0335-052E affords an opportunity to observe the process at reasonably
high resolution under conditions similar to star formation at very high
redshift.  There are several qualitative clues and a few quantitative
ones that can be identified.

\subsection{Qualitative Evidence} \label{ss-qual}

There is significant evidence as discussed by \citet{thu97} and now with
the increase of Pa$\alpha$ emission toward the southern clusters that there is
a progression of apparent ages, starting with the supernova cavity in the
northeast, ending with the obviously young S1 and S2 in southwest. This suggests
that each episode of star formation has positive feedback on subsequent star
formation.  The lack of a significant population of stars in the supernova
cavity suggests that the very first star formation episode ended very quickly,
blowing out the area but producing compression of the gas that lead to star
formation at S6, S5 and S4.  These then induced star formation at S3 and
subsequently at S1 and S2. More qualitative evidence of positive, or at least
nondestructive feed back is the double source, S3.  The two clusters are 
separated on the sky by only 26 pc and are therefore almost certainly associated.  
That two clusters can form stars in such close proximity is strong evidence that
adjacent star forming areas do not produce strong negative feedback. Although
not as intense as S1+S2, S3 has detectable Pa$\alpha$ emission indicating current
star formation.  It is not possible to determine which of the clusters, or
perhaps both, is the source of the star formation.  The exact nature of the
clusters is not critical, but rather that two clusters can form
in close proximity, regardless of the sequence, is evidence that the presence
of large scale star formation does not necessarily terminate other star formation
in the immediate environment. It is 
unfortunate that we can not observe this system at much higher resolution to see 
the dynamical interaction between two closely paired clusters.

The simple existence of SSCs, which have thousands of O stars within a
radius of only a few parsecs, sets severe limits on the negative feedback
from massive star formation.  The formation of the first massive stars
does not prevent the subsequent formation of thousands of more massive
stars at average distances of far less than a pc from the first stars.
In fact this is a motivating consideration in the construction of our
model system to see if there is a plausible model that matches the observed
parameters and that allows the formation of such a cluster.  Since a large 
fraction of stars may have
formed in clusters this has important implications on theories of galaxy 
evolution.

\subsection{Quenched Photoionization Feedback} \label{ss-quan}

If our model of individual HII regions surrounding the O stars is even
approximately correct it says that the volume of ionized gas in at least 
S1 is quite small compared to the volume of the cluster.  This invokes
a scenario that most of the luminous stars are losing mass through stellar 
winds and the ionizing photons of the stars are expended within a few 
100 AU from the stars due to the high densities created by the winds. 
This quenches any negative feedback due to the photoionization 
of neutral gas.  Other than through winds and gravity these stars have very 
little effect on their environments outside of their HII regions.  As long as 
these winds continue, photoionization will not play a role in feedback on the
galaxy as a whole. It is important to note here that the boundaries of the 
HII regions are simply where the ionizing photons are depleted and do not
necessarily mark any boundary for the gas outflow.

Although the effects of photoionization can be accurately mapped through 
emission line images and radio emission maps, the effects of stellar winds
are less easily measured.  Work by \citet{che85} indicated that massive and
concentrated star clusters should produce a strong stationary superwind at
all times in their evolution. This would remove gas from the cluster and
affect the surrounding region through shocks and density enhancement.  Recent
work by \citet{dop05}, however, indicates that many outflows around HII regions created
by young stellar clusters are "stalled" by the ambient interstellar medium.
In this case, however, the individual stellar HII regions appear to be stalled
by the intercluster gas.  This may be a phase that further delays the onset
of feedback at the cluster level.  \citet{dop05} note that it is mainly the
pressure that controls the expansion of the HII regions.  The low star to
gas mass ratio and the large gas mass of SBS0335-052 may contribute to
stalling out the expansion of the stellar HII regions.

\section{Future Evolution}

The future evolution of the SSCs and of SBS0335-052 itself is of great
interest.  SBS0335-052W has an even lower metallicity than the eastern
component but with a higher x-ray luminosity (\citet{thu05}, \citet{izo05}.  
It appears that the western component has not yet had a major episode of star
formation.

\subsection{Evolution of SBS0335-052}

As suggested by \citet{pus01}, the star formation in the eastern component
was probably triggered by interactions, either between the eastern and 
western component or with NGC 1376.  Since the two components share a 
common HI cloud it appears likely that it was the interaction with 
NGC 1376 that triggered the formation of the SSCs.  The large amount of
neutral gas, $2 \times 10^9$ M$_{\sun}$, shows that there is adequate mass
for further star formation in both components.  Based on only the radial component
of the velocity difference between SBS0335-052 and NGC 1376, they are
gravitationally bound to each other.  It would appear that the eventual fate
of SBS0335-052 is to merge with the much larger NGC 1376 in what in hierarchical
galaxy assembly would be called a minor merger.  In most galaxy merging
simulations significant star formation occurs after the first encounter
which would mean that SBS0335-052 has already had its first encounter with
NGC 1376 and may have been significantly closer to it in the past.

\subsection{Evolution of the SSCs}

The current SSCs in SBS0335-052E will evolve on a much shorter time scale
than the SBS0335-052/NGC 1376 system.  According the stellar populations
calculated by Starburst99 there are more than 2500 stars in the S1 
SSC with more than 30 M$_{\sun}$.  These stars should end up as black 
holes at the end of their evolution.  
Although many of the black holes may be ejected from the cluster due to 
supernova kicks, there exists the possibility of having several thousand
black holes within the 3 pc radius of the cluster.  This proximity may
lead to mergers which could form an intermediate mass black hole, consistent
with the x-ray emission from the region.  An alternative scenario is 
presented by \citet{zwa04} who modeled the cluster MGG 11 in M82.  They
calculated that the dynamical time for a 100 M$_{\sun}$ star to sink
to the center of the cluster was 3 Myr which led to several collisions
and the formation of a 800-3000 M$_{\sun}$ star which then led directly to
the creation of an
intermediate mass black hole without significant mass loss.  Although
the radius of S1 is about twice that of MGG 11 it has about 10 times
more mass leading to a similar dynamical sink time from equation 1 
of \citet{zwa04}.  In either scenario super star clusters may be the
birthplace of intermediate mass black holes that can be the seeds of
the massive black holes that power AGNs.  The lack of x-ray emission
from the other sources in SBS0335-052 indicate that this process did
not occur in those sources, however, we do not currently have detailed
information on their stellar density in the formation epoch equivalent
to the current epoch of S1 and S2.

\section{Conclusions}

The Pa $\alpha$ flux from the S1 SSC provides a determination of
the volume emission measure which when compared to the observed
radio emission indicates that much of the radio emission occurs
at high to moderate optical depth and high density.  A simple model
which divides the volume emission measure equally among the 7500 O
stars predicted from STARBURST99 and using radio
emission models of \citet{oln75} indicates a very low filling factor
for the ionized gas, consistent with the assumption of individual
HII regions.  The confinement of the ionizing flux to regions of
only a few hundred AU around individual stars limits and delays the
effect of photoionization of neutral gas and molecular clouds.  This
provides a consistent picture of how several thousand O stars can 
form within a radius of 3 pc and may be indicative of star formation
at early times in the universe. Although indicating very little
negative feedback at the current time, SBS0335-052 has a very low
star to gas mass ratio.  As that ratio increases and the gas mass
decreases the confining pressure of the interstellar medium may
decrease to a level where the HII regions are no longer stalled
at small radii and begin to produce negative star formation feedback
on the galaxy as a whole. Finally we address the evolution of both the 
complex and the SSCs indicating that SBS0335-052 may be part of a minor
merger with NCC 1376 and speculating that S1 and SSCs in general may
be the birthplace of intermediate mass black holes.

\acknowledgments

This work contains data from the NASA/ESA Hubble Space Telescope which is
operated by the Association of Universities for Research in Astronomy
(AURA) Inc. under NASA contract NAS5-26555. The introductory text benefited
from slight variations of text contributed by John Peacock for another
purpose.  RIT would like to thank Leslie Hunt, Chris McKee and Kelsey Johnson 
for very useful conversations.  We would like to acknowledge the useful
comments and suggestions from an anonymous referee.

\clearpage

\begin{deluxetable}{cccc}
\tabletypesize{\scriptsize}
\tablecaption{Pa$\alpha$ and Br$\alpha$ Line Fluxes \label{tbl-paba}}
\tablewidth{0pt}
\tablehead{
\colhead{Line} & \colhead{Observed\tablenotemark{a}}   & \colhead{Extinction Corrected}   &
\colhead{Predicted\tablenotemark{b}} \\
\colhead{ } & \colhead{erg s$^{-1}$ cm$^{-2}$} &  \colhead{erg s$^{-1}$ cm$^{-2}$} &
 \colhead{erg s$^{-1}$ cm$^{-2}$}
}
\startdata
Pa$\alpha$& $1.77 \times 10^{-14}$ & $8.01 \times 10^{-14}$ & $6.42 \times 10^{-14}$\\
Br$\alpha$& $9.0 \times 10^{-15}$ & $1.52 \times 10^{-14}$ & $1.52 \times 10^{-14}$
\enddata

\tablenotetext{a}{$1 \arcsec$ by $1.5 \arcsec$ aperture}
\tablenotetext{b}{For 20,000 K gas}

\end{deluxetable}

\clearpage

\clearpage

\begin{thebibliography}{}

\bibitem[Bauer et al.(2005)]{bau05} Bauer, A.E., Drory, N., Hill, G.J.
	\& Feulner, G. 2005 \apj, 621, L89

\bibitem[Benson et al.(2003)]{ben03} Benson, A.J., et al. 2003, \apj, 599, 38

\bibitem[Chevalier \& Clegg(1985)]{che85} Chevalier, T.A., \& Clegg, A.W.
	1985, \nat, 317, 44

\bibitem[Cole et al.(2001)]{col01} Cole, S., et al. 2001. \mnras, 326, 255

\bibitem[Dale et al.(2001)]{dal01} Dale, D.A. et al. 2001, \aj, 122, 1736

\bibitem[Dopita et al.(2005)]{dop05} Dopita, M.A. et al. 2005, \apj, 619, 755

\bibitem[Engelbracht et al.(2005)]{eng05} Engelbracht, C.W. et al. 2005
	in ASP Conference Series TBD, The Spitzer Space Telescope:
	New Views of the Cosmos, L. Armus ed.  in press

\bibitem[Houck et al.(2004)]{hou04} Houck, J.R. et al. 2004, \apjs,
	154, 211

\bibitem[Hunt et al.(2004)]{hun04} Hunt, L.K., Dyer, K.K., Thuan, T.X.
	\& Ulvestad, J.S. 2004, \apj, 606, 853

\bibitem[Hunt, Vanzi \& Thuan(2001)]{hun01} Hunt, L.K., Vanzi, L.,
	\& Thuan, T.X. 2001, \aap, 377, 66

\bibitem[Izotov et al.(1990)]{iz90} Izotov, Y.I. et al. 1990, \nat, 343, 238

\bibitem[Izotov et al.(1992)]{iz92} Izotov, Y.I., Lipovetsky, V.A., Guseva, 
	N.G., \& Kniazev, A.Y. 1992, in The Feedback of Chemical
	Evolution on the Stellar Content of Galaxies, ed. 
	D. Alloin \& G. Stasinska (Paris: Obs. Paris), 138

\bibitem[Izotov et al.(1997)]{izo97} Izotov, Y.I. et al. 1997, \apj, 476, 698

\bibitem[Izotov, Thuan \& Guseva(2005)]{izo05} Izotov, Y.I., Thuan, T.X.,
	\& Guseva, N.G. 2005, \apj, in press. 

\bibitem[Johnson(2005)]{joh05} Johnson, K. 2005, IAU Symposium 227,
	online record of presentations, http://www.arcetri.astro.it/iaus227/,
	slides 8 and 10 in talk, (click on talk not the presentation title).

\bibitem[Kennicutt(1998)]{ken98} Kennicutt, R.C. 1998, \araa, 36, 189

\bibitem[Kennicutt et al.(2003)]{ken03} Kennicutt, R.C. et al. 2003, 
	\pasp, 115, 928

\bibitem[Kunth et al.(2003)]{kun03} Kunth, D. et al. 2003, \apj, 597, 263.

\bibitem[Kunth et al.(2004)]{kun04} Kunth, D. et al. 2004,
	astro-ph/0407584 v1

\bibitem[Leitherer et al.(1999)]{lei99} Leitherer et al. 1999, \apjs, 123, 3

\bibitem[Lytle et al.(1999)]{lyt99} Lytle, D., Stobie, E., Ferro, A.,
	\& Barg, I. 1999, in ASP Conf. Ser. 172, Astronomical Data
	Analysis Software and Systems VIII, ed. D. Mheringer, R. 
	Plante, \& D. Roberts (San Francisco; ASP), 445

\bibitem[Markarian, Lipovetsky \& Stepanian(1983)]{mar83} Markarian,
	B.E., Lipovetsky, V.A \& Stepanian, J.A. 1983, Astrofizika,
	19, 29

\bibitem[Olnon(1975)]{oln75} Olnon, F. M. 1975, \aap, 39, 217

\bibitem[\"{O}stlin \& Kunth(2001)]{ost01} \"{O}stlin, G. \& Kunth, D. 2001,
	\aap, 371, 429

\bibitem[Papaderos et al.(1998)]{pap98} Papaderos, P., Izotov, Y.I.,
	Fricke, K.J., Thuan, T.X. \& Guseva, N.G. 1998, \aap, 338, 43

\bibitem[Plante \& Sauvage(2002)]{pla02} Plante, S. \& Sauvage, M. 2002,
	\aj, 124, 1995

\bibitem[Pustilnik et al.(2001)]{pus01} Pustilnik, S.A., Brinks, E.,
	Thuan, T.X., Lipovetsky, V.A. \& Izotov, Y.I. 2001, \aj,
	121, 1413

\bibitem[Pustilnik, Pramskij, \& Kniazev(2004)]{pus04} Pustilnik, S.A.,
	Pramskij, A.G. \& Kniazev, A.Y. 2004, \aap, 425, 51

\bibitem[Rieke \& Lebofsky(1985)]{rie85} Rieke, G.H. \& Lebofsky, M.J.
	1985, \apj, 288, 618

\bibitem[Schaerer(2002)]{sch02} Schaerer, D. 2002, \aap, 382, 28

\bibitem[Simon et al.(1983)]{sim83} Simon, M., et al. 1983, \apj, 266, 623

\bibitem[Sternberg, Hoffmann \& Pauldrach(2003)]{str03} Sternberg, A.,
	Hoffmann, T.L. \& Pauldrach, A.W.A. 2003, \apj, 599, 1333

\bibitem[Thompson(1982)]{thm82} Thompson, R.I. 1982, \apj, 257, 171

\bibitem[Thompson(1984)]{thm84} Thompson, R.I. 1984, \apj, 283, 618

\bibitem[Thompson et al.(2005)]{thm05} Thompson, R.I. et al. 2005,
	\aj, 130, 1

\bibitem[Thuan, Izotov \& Lipovetsky(1997)]{thu97} Thuan, T.X., Izotov,
	Y.I., \& Lipovetsky, V.A. 1997, \apj, 477, 661

\bibitem[Thuan \& Izotov(1997)]{thu97b} Thuan, T.X. \& Izotov, Y.I. 1997, \apj, 489, 623

\bibitem[Thuan, Sauvage \& Madden(1999)]{thu99} Thuan, T.X., Sauvage, M.,
	\& Madden, S. 1999, \apj, 516, 783

\bibitem[Thuan et al.(2004)]{thu04} Thuan, T.X., Bauer, F.E., Papaderos, P.,
	\& Izotov, Y.I. 2004, \apj, 606, 213

\bibitem[Thuan, des Etangs \& Izotov(2005)]{thu05} Thuan, T.X., 
	des Etangs, A.L, \& Izotov, Y.I. 2005, \apj, 621, 269

\bibitem[Vanzi et al.(2000)]{van00} Vanzi, L., Hunt, L.K., Thuan, T.X.
	\& Izotov, Y.I. 2000, \aap, 363, 493

\bibitem[Woan(2000)]{wo00} Woan, G. 2000, \emph{The Cambridge Handbook
	of Physics Formulas}, (Cambridge, England, Cambridge University 
	Press), p160.

\bibitem[Zwart et al.(2004)]{zwa04} Zwart, S.F.P., Baumgardt, H., Hut, P.,
	Makino, J. \& McMillan, L.W. 2004, \nat, 428, 724

\end{thebibliography}
\end{document}